# Rational AI: A comparison of human and AI responses to triggers of economic irrationality in poker

*Working paper*

November 2021

C. Grace Haaf, Department of Business and Finance, New York University Shanghai
Devansh Singh, ZS Consulting
Cinny Lin, New York University Shanghai
Scofield Zou, New York University Shanghai

## Abstract

Humans exhibit irrational decision-making patterns in response to environmental triggers, such as experiencing an economic loss or gain. In this paper we investigate whether algorithms exhibit the same behavior by examining the observed decisions and latent risk and rationality parameters estimated by a random utility model with constant relative risk-aversion utility function. We use a dataset consisting of 10,000 hands of poker played by Pluribus, the first algorithm in the world to beat professional human players and find (1) Pluribus does shift its playing style in response to economic losses and gains, *ceteris paribus*; (2) Pluribus becomes more risk-averse and rational following a trigger but the humans become more risk-seeking and irrational; (3) the difference in playing styles between Pluribus and the humans on the dimensions of risk-aversion and rationality are particularly differentiable when both have experienced a trigger. This provides support that decision-making patterns could be used as "behavioral signatures" to identify human vs. algorithmic decision-makers in unlabeled contexts.

## Introduction

Humans react emotionally to economic losses and gains, for example, becoming more risk-averse as the result of fear or more risk-seeking and impulsive while in a positive emotional state [1], [2][1]. Artificial intelligence (AI) will not be subject to *emotionally-driven* changes in its decision-making process. Its decisions are governed by mathematical optimization of a pre-determined objective function. While one could specify an objective function that synthetically

---

[1] Reactions are heterogeneous, i.e. not all humans become more risk-averse as a result of experiencing fear or risk-seeking when in a positive emotional state

mimics the effects of emotion, the bits and bytes are not experiencing emotion as humans understand it.

In many contexts, particularly commercial environments, we would expect that AI governing functions optimize for traditional economic objectives, e.g. maximize profit or minimize potential loss. If we define rational behavior as making decisions that are consistent and coherent with maximizing utility [3], then an AI would be expected to behave rationally given its required adherence to optimizing the objective function. Furthermore, we would not expect utility functions to reflect non-economic concepts related to emotion—for example, proxying human ego by explicitly maximizing the pleasure derived from being held in high-esteem by peers—even if it is possible to synthetically imitate this concept[2]. With the goal of investigating how humans and AI differ on the "how" and "why" of departures from economic rationality, this paper compares and contrasts how the two react to economic triggers like losses and gains in a game of poker.

We use data from poker games that consist of humans and an AI, Pluribus. Pluribus' predecessor, Liberatus, was the first AI to beat professional human players in a heads-up (two-player), no-limit poker game [4], and Pluribus was the first AI to beat the professionals in a six-person, no-limit tournament format in 2019 [5]. As discussed in Brown and Sandholm [5], Pluribus played a series of games against the professional human poker players prior to the final tournament in order to train its model. We use this publicly available training data for our analysis. Poker – and this dataset in particular – are apt for this investigation because:

- Objectives in poker and other forms of structured gambling have straightforward economic objectives. Presumably all players, human and algorithmic, have a singular, identical goal, which is to win money by winning poker hands.
- The decision space, or actions available to players, is finite and small. Players can choose to play or fold in any given turn. The only continuous variable is bet amount, and that is bounded by the number of chips a player has and other players' previous actions. In this no-limit game the maximum bet is $10,000 and bets are placed on a continuous scale. Pluribus is constrained to a few bets between $100 and $10,000 by its specification, but the humans are not [5].
- This particular dataset has a sufficient number of observations per individual to fit models with statistical significance (9,933 choice situations for Pluribus and 49,665 for the humans)
- In poker contexts, humans are known to exhibit tilt: "the mental or emotional confusion or frustration in which a player adopts a less than optimal strategy, usually resulting in the player becoming over-aggressive" [6].

We define a trigger as situation in which humans often experience a change in their decision-making process, presumably partially driven by emotion. For our poker dataset we examine the effect of the trigger situations "experiencing a loss" and "experiencing a win" on players' decision to play or fold in the pre-flop round of the subsequent hand. Our hypothesis is that following a trigger event, a human player will exhibit a change in playing style, becoming more or less rational. The algorithmic agent, Pluribus, will not.

---

[2] Despite these expectations, AI may exhibit behavior that appears irrational, but for reasons other than emotion: mathematical approximations used in the objective function, computational errors like spurious optimizers, or bugs in the code, to conjecture a few.

- **Research question 1: does an algorithm change its playing style as a result of a trigger?**
- **Research question 2: how does playing style change?**
- **Research question 3: why do playing styles change?**

## Literature Review

Tversky and Kahneman [3] characterize rational behavior as consistent and coherent. Normative economic models—models that prescribe what an agent *should* choose when presented a set of probabilistic, monetary outcomes— by and large do not match what agents actually choose, which would lead to the conclusion that those agents are irrational [7]. Descriptive models that form the basis of descriptive decision theory seek to explain why the agents made the observed choices that they did, assuming that their choices are mostly rational but subject to consistent (unobserved) rules, which affect the evaluation of their observed choices as coherent and consistent.

**Towards coherence**: Tversky and Kahneman [7] argue that "the deviations of actual behavior from the normative model are too widespread to be ignored, too systematic to be dismissed as random error, and too fundamental to be accommodated by relaxing the normative system". In contrast, they illustrate that models that account for individual risk attitudes do not violate fundamental assumptions of rationality when applied to observed decisions[3]. Theoretically, these models will be able to differentiate between actual irrational decisions and spurious effects due to omission of the risk tolerance explanatory variable present in normative models[4].

**Towards consistency**: Even under more coherent models that represent individual risk attitudes, we may still observe inconsistencies (and incoherence) over a set of actual decisions, i.e. agents will make different choices for the same set of options and outcomes based on a change in framing [3]. For example, in an experiment, participants reversed their choice of a preferred crisis-response program based on whether the program outcomes were framed as loss aversion (death) versus potential upside (lives saved), even though the number of probabilistic deaths was equivalent. It is debatable whether inconsistencies due to framing are irrational[5]. In this paper, our choice situation option and outcome framing remain constant across observed decisions: players are only allowed to play or fold, the outcomes are win or lose, and the rules of the game do not change.

---

[3] Specifically, Kahneman and Tversky [7] illustrate that prospect theory, a specific type of descriptive decision model, is the only model as compared to various formulations of normative models that does not violate assumptions on dominance (contrasting risk attitudes) and invariance (framing effects) across choice situations.

[4] The model may still detect spurious effects for many reasons, including modeling errors related to risk tolerance (no model is perfect), but the model is at least partially controlling for this factor.

[5] For example, Steverson et al. [14] propose that human decision-making is influenced by neurobiological factors, such as reducing computational overhead at the expense of precision in decision-making, i.e. more observed inconsistency. Particular types of framing may carry more computational load or thermodynamic cost for the brain, and humans may be jointly optimizing their choices to reduce this load in conjunction with maximizing utility from other sources.

In this paper, we compare and contrast human and AI choices in a game of poker on the dimensions of coherence and consistency to examine the agents' relative rationality. We investigate the effect of prior outcomes—wins or losses—on estimated risk-aversion (and irrationality) in subsequent choice situations. Similar to Thaler and Johnson [8], we find that prior decision outcomes affect future risk tolerance. Smith et al [9] and Eil and Lien [10] investigate the effect of wins and losses in poker on humans' risk-aversion. Both papers use data from online poker site, Full Tilt, to analyze experienced poker players' decisions. Smith et al. [9] find that players become more "loose" (play a greater percentage of hands voluntarily) following a loss than following a win and hypothesize that this is due to the "break-even" effect prevailing over the "house-money" effect, coherent with Thaler and Johnson (1990). Following a big win, the players become less aggressive[6] as compared to following a loss. Eil and Lien [10] examine the adherence of experienced poker players to expected utility and prospect theory models. They find evidence of the "break-even" effect but that experienced players do not exhibit the "house money" effect (wins make them more risk-averse).

In this paper, we analyze poker decisions by estimating a random utility model (RUM) with the risk tolerance parameter entering through the utility function. Smith et al [9] estimate a Wilcoxon signed-rank test for paired differences on the calculated looseness and aggression statistics for players following wins and losses. Eil and Lien [10] estimate a Cox proportional hazard model on covariates for win and loss to predict the probability that a player will quit an online poker session. Under the supposition that experiencing either a win or loss will increase risk-aversion according to cumulative prospect theory, they expect a win/loss to increase the propensity that a player will quit the game.

Intuitively, we expect humans to alter their attitudes towards risk due at least in part to emotional factors. Heilman et al [11] find that naturally occurring negative feelings increase risk-aversion but emotional regulation, or reappraising negative emotions, reduces it. Nguyen and Noussair [12] find a more positive emotional state leads to more risk-taking, but stronger emotions—including positive ones like happiness—result in more risk-aversion. Campos-Vasquez and Cuilty [13] find risk-aversion increases with sadness, and anger reduces loss aversion by half. Coget et al. [14] conclude anger and fear influence rational decision-making, though how they do so is contingent on other internal and external factors such as degree of prior experience with the choice situation. Schulreich et al [15] induce fear before or during decision-making and find a positive correlation between being in a fearful state and exhibiting risk-aversion when choosing between mixed gambles. Cristofaro [16] contains a meta-analysis of 123 studies that investigate how emotion impacts economic decisions in managerial contexts. While we do not *explicitly* investigate the role of emotion in this paper, the potential for the presence of human emotion in decision-making is a key driver leading to our hypothesis

---

[6] Aggressiveness is defined the ratio of check and call actions to bet and raise actions

that Pluribus and the humans will exhibit differentiable risk tolerances and reactions to losses and wins, since an AI agent will definitely not experience these emotions[7].

Research on AI decision-making to date has focused on how to communicate AI decisions to humans. Hind et al [17] propose a framework for simple communication of AI process and output. They include a literature review that divides related work on explaining AI decisions into three categories: (1) technical model interpretation, (2) creating simplified models in the explanatory stage, and (3) natural language processing and computer vision that derives rationales from input text. Research at the intersection of AI and irrationality is concentrated primarily in the domain of computer science and specifically as a topic in inverse reinforcement learning. This subdomain focuses on how to adapt policy learning functions to account for known irrationality in humans [18].

To the authors' knowledge, this paper is the first to apply descriptive decision theory models to examine the potential irrationality in AI decision-making through the lens of economic rationality on the basis of consistency and coherence.

## Methods

### Data overview

We use the published Pluribus training dataset available from Brown and Sandholm [5]. The AI, Pluribus, plays a series of Texas Hold 'Em poker games, each against five human opponents. There are a total of 67 games played by Pluribus and 13 different human opponents[8]. Pluribus plays in every game. We pool the data for the humans to proxy average human behavior, we term this pooled agent "Human", using the plural personal pronouns they/their/them to emphasize that it is comprised of multiple agents. A game ends after a specified amount of time has elapsed and on average, there are 149 hands played per game. Players are incentivized with monetary compensation; however, the winnings are not directly paid to the participants, and at the beginning of each hand, the amount each player has available for wagering (stack size) is reset.

Games are subdivided into hands, which can be additionally subdivided into rounds of betting. A player wins a hand by (a) having the best set of five cards possible as compared to other players at the end of the hand or (b) forcing all other players to quit the hand ("fold") before all betting rounds have been completed rather than continue to put more money at risk. In the first round of each hand ("pre-flop"), two cards are dealt to each player face-down (the other players do not know the value of these cards). In subsequent rounds, additional cards will be dealt face-up and shared by all players. The first two players to the left of the dealer, the "small blind" and the "big blind" are required to commit at least a minimum amount of money to the pot in the pre-flop round. The deal rotates each hand so that a player is in these roles an equal number of times.

---

[7] While an AI could hypothetically be specified to include a synthetic approximation of emotion, Pluribus does not explicitly (or intentionally implicitly) seek to represent it.

[8] There are 13 players named by an alias in the dataset (excluding Pluribus), however, it appears that there may be mislabeled data. See Appendix C for details on name discrepancies.

## Models overview

We want to understand if the AI changes its playing style in regards to rationality as a result of a trigger, and if so, how this compares to the behavior of humans. We characterize this behavior on two dimensions: (1) observed outcomes and (2) model-estimated latent decision-making parameters of the players.

We define a trigger as an event that causes humans to shift their approach to risk-taking holding the observed, statistical drivers of the decision-making context constant. For example, if a player were perfectly rational and had reached a steady-state calibration for perceived risk and reward (playing more hands does not cause a player to update their beliefs about the game or their opponents), then they should always make the same decision about playing or folding a hand if all other factors are held constant. The decision to play (or fold) a hand should not be affected by the outcome of the previous hand. The potential trigger will then be an experienced win- or loss-outcome in the previous hand.

In order to limit the complexity of modeling joint player actions, we only consider players' decisions pre-flop. Considering the pre-flop decision has the additional advantage that the decision is occurring as close in time to the loss or win itself, so we would expect the emotional impact to be strongest since the trigger is still fresh in the player's mind.

## Random Utility Model

We specify RUM:

$$\hat{y}_{in} = \frac{e^{\lambda_i u_{inj}}}{\sum_{k \in play, fold} e^{\lambda_i u_{ink}}} \qquad (1)$$

where $\hat{y}_i$ is the predicted probability that a player $i$ will choose option $j$ for a given choice situation (hand) $n$ and $\lambda$ is the precision parameter or rationality parameter [13]. The utility of the options "play" or "fold" is evaluated using constant relative risk-aversion (CRRA):

$$u_{inj} = \sum_{o \in win, lose} p_{on} \frac{v_{on}^{1-\omega_i}}{1-\omega_i} \qquad (2)$$

where $o$ indexes outcomes "win" and lose", $p$ and $v$ are the probabilities and payoffs associated without the outcome, and $\omega$ is a player's risk-aversion level. We fit a model to the set of observed choices, $\mathbf{y}_i$, to estimate a player's risk-aversion, $\omega$, and rationality, $\lambda$, using maximum likelihood estimation. We partition a player's data in to three sets: hands played following a loss in the previous hand, following a win, and following a neutral-outcome. To control for the influence of trivial outcomes that are based only on the (required) blinds we define a win as a pot containing more than just the small and big blinds, a loss as losing more than the amount of the big blind, and a neutral outcome as all other outcomes (blinds only). We can then compare the values of $\omega$ and $\lambda$ to investigate how a player's risk-aversion and rationality change in the subsequent decision as a result of experiencing a trigger.

**Estimation:**

For all choice situations in our model, the option "play" is the riskier option, meaning that only for low levels of risk-aversion do players prefer "play" to "fold". There are three types of gambles faced in our dataset:

- **risk-dominant**: for all levels of risk-aversion, $\omega$, the riskier option has a greater CRRA utility than the safer option
- **safe-dominant**: for all levels of risk-aversion, $\omega$, the safer option has a greater CRRA utility than the risky option
- **mixed**: for some levels of risk-aversion the riskier option is preferred and vice-versa

The majority of gambles in the dataset are safe-dominant and the players choose to fold (Table 1). For these types of gambles and choices, the likelihood of the model is optimized as $\omega \rightarrow -\infty$ (this is also the case for risk-dominant gambles when players choose to play, see Figure 1(a, b)).

Apesteguia and Ballester [19] investigate the drawbacks of the RUM-CRRA specification and primarily raise the concern that there is an identification issue with $\omega$ due to non-monotonicity of the RUM in $\omega$ (Figure 1c). For some levels of $\omega$, the relative attractiveness of the risky option increases with increasing risk-aversion. We verify that the expected $\omega$ calculated from the results of the valid convergences of multi-start optimization is in the allowable domain for almost all choice situations (<15%), i.e. that it is in the domain of $\omega$ for which the relative attractiveness of the risky option *decreases* with increasing risk-aversion.

For ease of computation, all payoffs, $v$, are shifted to be in the positive domain and such that the CRRA utility is roughly $O(1)$. The payoffs were mapped to a standard normal distribution and then shifted by a constant such that the smallest payoff in all choice situations for all players was located at a value of 1.

*Table 1 – Player observed choice by CRRA gamble type*

|  | Mixed gamble | | Risky gamble dominates for all risk-aversion levels | | Safe gamble dominates for all risk-aversion levels | |
|---|---|---|---|---|---|---|
| Player | fold | play | fold | play | fold | play |
| Pluribus | 16% | 5% | 12% | 11% | 46% | 10% |
| Human | 17% | 6% | 8% | 9% | 48% | 12% |

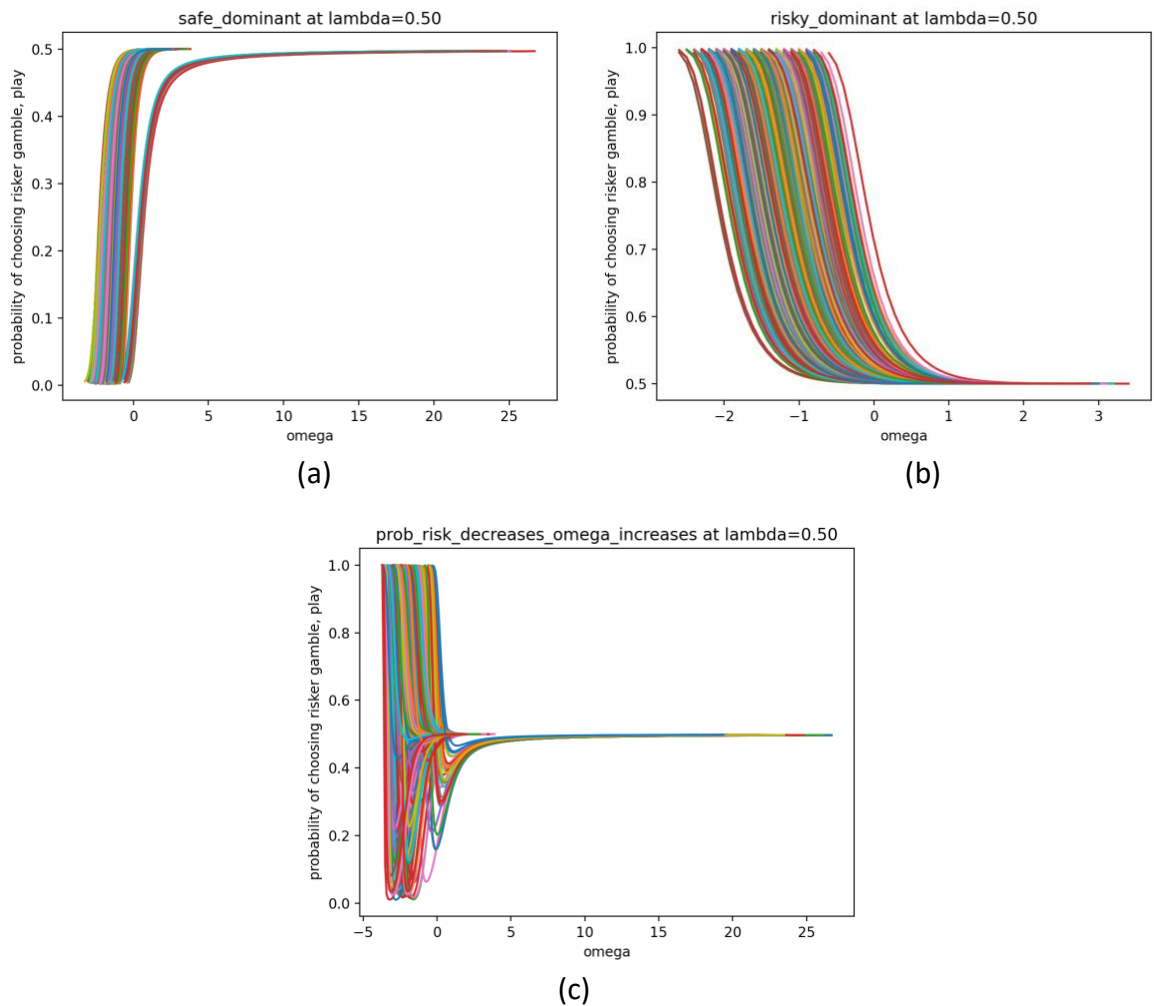

*Figure 1 – Probability of choosing riskier option (play) by risk-aversion level for choice situations faced by Pluribus in a neutral state. Increases/decreases are monotonic in risk-aversion for safe-dominant (a) and risk-dominant (b) gambles, but non-monotonic for mixed gambles (c).*

## Deep Neural Net

Estimation of the RUM is dependent on the probabilities and payoffs of the gambles. In Texas Hold 'Em, the payoffs occur at the end of the hand (player wins or loses), *after* the board cards have been dealt. The choice we are modeling—the decision to fold in the pre-flop round—occurs *before* the board cards have been dealt. The players are weighing the probability that they will ultimately win the hand given what private cards they are holding, what their opponents are holding, and what will be revealed when the subsequent board cards are dealt. They not only have to decide whether they want to fold, accepting a loss (or at least not a win), but also how much money to put at risk and forecasting what others will choose to risk.

It is currently intractable to calculate the actual probability of winning the hand given the cards in one's hand (and in fact why it is only recently that a computer could beat humans! [5]). We estimate two sub-models to approximate the probabilities and payoffs that the players are weighing when making their pre-flop decisions using the dataset itself. Dependent upon how strong a player's priors on probability and payoffs are, this method may or may not most closely match the player's **perception**, on which they are basing their choices. If players index heavily on context-specific information and dynamically update their perceptions during play, this method would closely match the actual situations the players are facing, i.e. their experiences with this particular set of opponents and styles. However, if players lean on external-context priors, this particular sample of data may diverge from their perceptions such that the model is a poor representation of their choice framework.

**Deep neural net for forecasting probability of winning**

We fit a deep neural net (DNN) to all hands in the dataset to predict the probability of binary outcome variable "win hand" using the features in Table 2. Stack size was not included since in this game format the player stacks were reset after each hand. The model obtained 58% accuracy on a test dataset.

*Table 2 – Deep neural net features used for predicting hand outcome*

| Feature | Description | Representation of decision-making factors |
|---------|-------------|-------------------------------------------|
| Sklanksy rank | Ranges from 1 (strongest hand) to 9 (weakest hand) | Strength of hole cards (two pre-flop private cards) |
| Seat | • Number of places to left of dealer, ranges from 1-6 in our dataset<br>• Seat 1 is on the dealer's immediate left and is the small blind and seat 2 is the big blind | • Amount of information when making decision due to previous player decisions<br>• Required commitment of money due to blinds |
| Player and seat | Dummy variable for given player in a given seat | Player response to seat |
| Sklansky and seat | Dummy variable for given Sklansky rank dealt to a given seat | Relative strength of cards given seat |

**Linear regression for forecasting winnings**

For the fold option, the player loses any contributions they had made to the pot in the pre-flop round. Note that folding in a pre-flop round is equivalent to losing with probability one. For the play option we fit two linear regressions, one for each of the possible outcomes:

- Win: regress the amount won on the pre-flop total amount of money in the pot and the number of participants choosing not to fold pre-flop ($R^2$ = 46.8%)
- Lose: regress the amount lost on individual player's pre-flop pot contribution ($R^2$ = 31.4%)

Post-estimation, the DNN and payoff models were applied to each observation (player choice situation) in the dataset to predict the probability of winning a hand as well as the associated win- and lose-payoffs. Given that the explanatory power of the DNN and linear regression models are low, there is cause for concern that interpretations and conclusions drawn from the risk and rationality parameters estimated by the RUM, which depend on the output of the DNN and payoff models, are invalid. However, referring to the definition of rationality in terms of coherence and consistency of decision evaluation discussed in the literature review, we would expect the results of the RUM to be at least directionally interpretable because any bias or error present in the DNN or logistic regression models will be consistent across framing of all choice sets presented to the RUM. This would not hold if there were systematic deviations in the distributions of choice framing (probabilities and payoffs) across post-loss, post-win, or neutral choice situations.

Additionally, there is the potential for significant and misleading coefficient bias in the RUM model parameters if the unexplained portion of probabilities or payoffs from the DNN or regression models are *sufficiently correlated* with an individual's *perception* of risk or reward in the choice situation (omitted variable endogeneity bias). This correlation may be present—and it is impossible to verify—but because we have controlled for the obvious factors available to participants in their evaluation of the probability and payoff of winning or losing (card strength, seat at the table, other players and their bets) we would not expect that the players' evaluation of outcome uncertainty is systematically related to the unexplained drivers of the outcome uncertainty; in fact, we expect that most of the *actual* outcome variation and players' *perceived* outcome variation is due to the number of possible permutations of cards dealt after the flop.

## Results

We characterize Pluribus' and the Human's behavior on two dimensions: the **observed** decisions made by the players in the game (play or fold) and the **latent** risk tolerance and rationality parameters estimated by the random utility model. The key difference between the two is that the model considers the relative quality of each decision faced, e.g. how good the player's hand is, whereas, the observed in-game decisions reflect the aggregate rationality of the set of decisions taken together. Loosely, the model parameters describe a player's internal decision-making process (coherence) and the set of observed decisions are how that process appears to observers in the external, real-world context (consistency).

For ease of discussion, we define the following terms in the context of this paper:

**"Should"**: we use this to refer to the play/fold decision a player would make if they were adhering to the estimated model recommendation, i.e. if the estimated utility of the option to play is greater than the estimated utility of the fold option, then a player *should* choose to play. This term is subjective generally, and even in our defined context it is still subjective. The estimated option utilities are subject to the uncertainty from (a) the DNN used to estimate outcome probabilities of winning/losing the hand and (b) the estimated risk-aversion parameter in the RUM (CRRA utility). In Figure 4, we show the *expected* utilities in order to illustrate the sensitivity (or lack thereof) to the RUM estimation.

**"Rational"**: the decision that a player *should* make in a given choice situation or, in aggregate, the proportion of decisions that agree with those indicated by the RUM, i.e. fold a hand with a relatively higher fold utility and play a hand with relatively higher play utility.

**"Triggered"**: we use this to refer to a player's hypothesized shift in mental state in the current choice situation they are facing following a win or a loss in the previous hand. We call these states "neutral", "post-loss", and "post-win", where post-loss and post-win are triggered states.

The players may or may not experience a shift in mental state due to having lost or won the previous hand—in fact, this is what we are testing—and we cannot say *what* would cause this shift in mental state if it exists. For the Human, we may be tempted to ascribe the shift to emotion, but we cannot actually know from this dataset. For Pluribus, emotions like ego or fear are not intrinsically present, and even terming a shift in decision-making behavior following a loss or win as a shift in "mental state" is probably slightly misrepresentative of what amounts to mathematical specification of an algorithm.

## Rationality of observed outcomes

Pluribus is biased towards folding hands. For situations in which the risky option (play) has a greater CRRA utility than the safe option (fold) regardless of a player's risk tolerance level, it still folds more than 50% of the time. The Human folds no more than 50% of the time. This assessment is subject to uncertainty from the DNN model but not the RUM (see Figure 2).

We expect that Pluribus and the Human will be better (more rational) at assessing option tradeoffs when one option is clearly better than the other—for example, playing a hand when the pre-flop cards are strong ("pocket aces") and folding when the cards are terrible (2 and 7 off-suit)—than when the relative option values are more uncertain, and in fact this is true (Figure 3). However, not only are Pluribus and the Human worse (more irrational) at assessing highly uncertain trade-offs, but both behave more irrationally when the tradeoff is irrational *and* the risky option is more attractive than the safe option. The more uncertain the environment, the more likely a player is to bias toward the safe option. In Figure 4, a greater proportion of irrational decisions (red-shaded bars) lie in the positive domain of the *x*-axis, which represents the difference in option utilities, and closer to the origin: these are decisions for which the risky option has a greater utility than the safe option but only marginally. Assessment of rationality in Figure 4 is dependent on the estimated probabilities and payoffs of the DNN model, but we show the expected values to mitigate the effects of RUM model parameter uncertainty, though the result is similar.

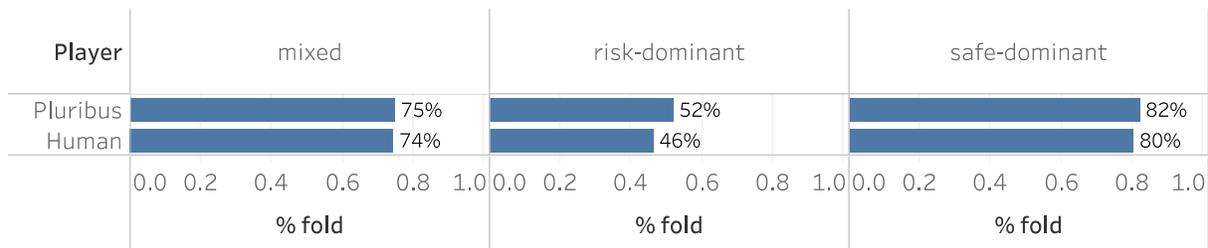

*Figure 2 – Pluribus is more likely to fold than the Human, particularly in low-uncertainty situations (risk-/safe-dominant CRRA gambles)*

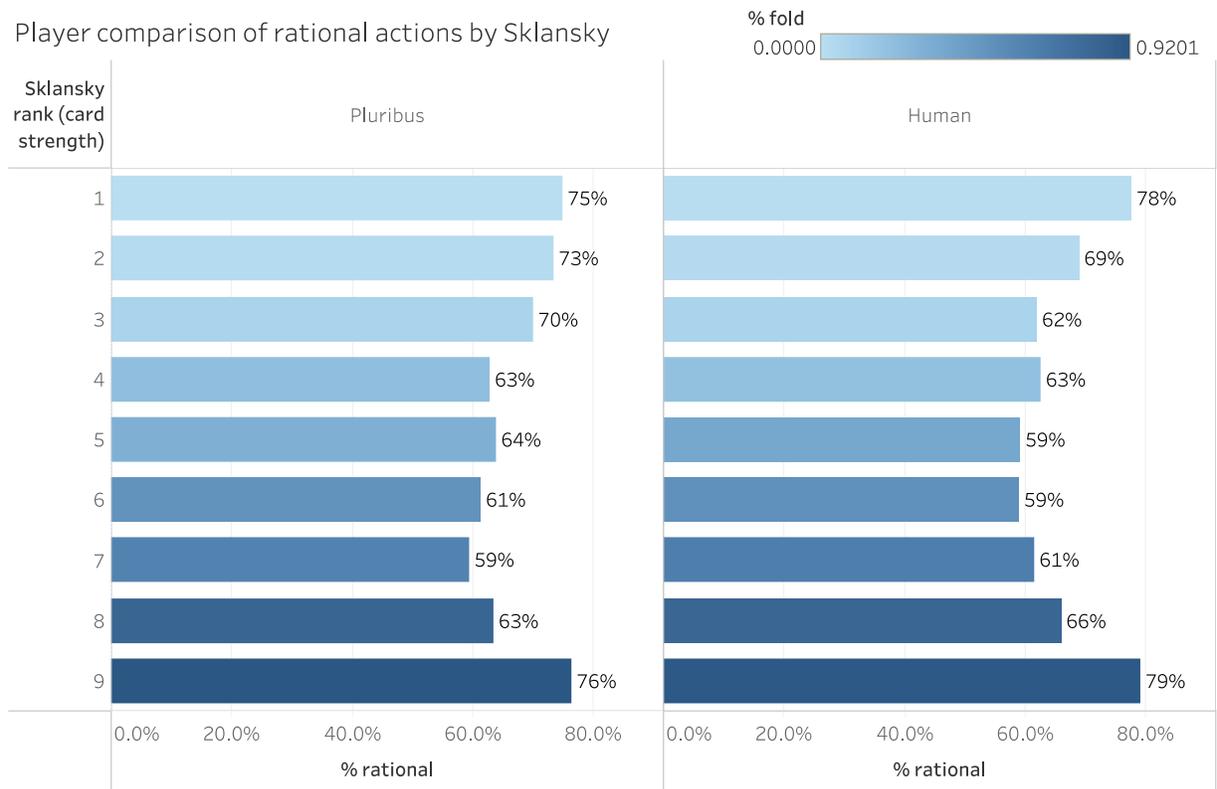

*Figure 3 – Pluribus and the Human make a greater proportion of irrational decisions in high-uncertainty environments (winning is less correlated with exogenous card strength)*

**Note: Sklansky rank ranges from 1=strongest cards through 9=weakest cards**

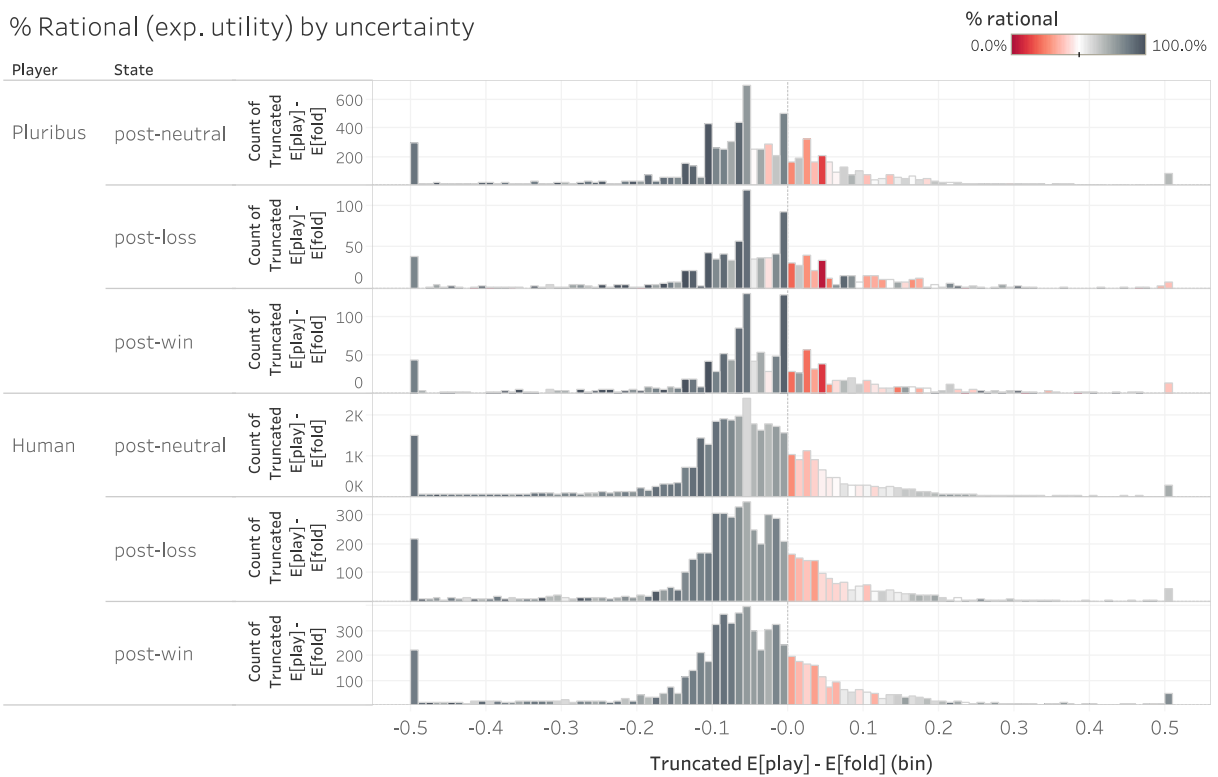

*Figure 4 – Comparison of rationality by relative utility difference of gambles, calculated using expected utility*

*Note: absolute utility differences > 0.5 mapped to point mass at +/- 0.5 for visualization purposes*

## Trigger effects

When Pluribus and the Human are triggered (in a post-loss or post-win state) they fold more than in the neutral-state (Figure 5), leading to an overall increase in observed rational decisions (Figure 7). This comparison holds for the risky- and safe-dominant gambles when subdivided by gamble type, and in more uncertain environments, i.e. the mixed-gamble choice situations, we see an even greater increase in the proportion of hands Pluribus folds. The Human does the opposite – they maintain their relative fold proportion in the neutral state when faced with uncertainty in a triggered state. A player's proportion of folds will be influenced by exogenous "luck-of-the-draw" factors (the strength of cards the player was dealt), but we do not see any meaningful variation in distribution of gamble quality across the neutral and triggered states in Figure 4, so this heuristic between-state comparison is weakly informative.

**Low-uncertainty environments (safe- or risk-dominant gambles)**: When Pluribus increases its fold percentage in the triggered states it only leads to an observed (statistically significant) change in rationality for post-win, risky-dominant gambles. Pluribus irrationally folds a greater proportion of hands. The Human, in contrast, does exhibit a statistically significant shift in rationality as a result of an increased proportion of folds. In safe-dominant gambles this equates to making more *rational* decisions, i.e. one should choose the safe option, fold. In risk-dominant gambles, this equates to making more *irrational* decisions, i.e. one should choose the risky option, play. Across all gambles, the increase in rationality prevails since the overall number of risk-dominant gambles is far fewer than the number of safe-dominant or mixed (one is far less likely to be dealt a strong hand than not). See Figure 8.

**High-uncertainty environments (mixed gambles)**: In mixed gambles where the assessment of what one *should* do is more uncertain and furthermore depends upon one's risk tolerance level, we see a divergence between the rationality shifts in Pluribus and the Human. For Pluribus, an increase in the proportion of folds results in a statistically significant increase in rational decisions because it takes fewer "foolish" chances: the proportion of irrationally-played hands decreases and the proportion of rationally-folded hands increases.

The Human maintains their overall proportion of play and fold as well as the implied level of rationality (the slight increase in rationality is not statistically significant). There are two factors that affect the perception of aggregate observed rationality under triggers: (1) by luck-of-the-draw, in the triggered situations, the Human faced relatively more choices for which the risky gamble (play) was rational but (2) they are still misidentifying hands that should be folded. See Figure 9.

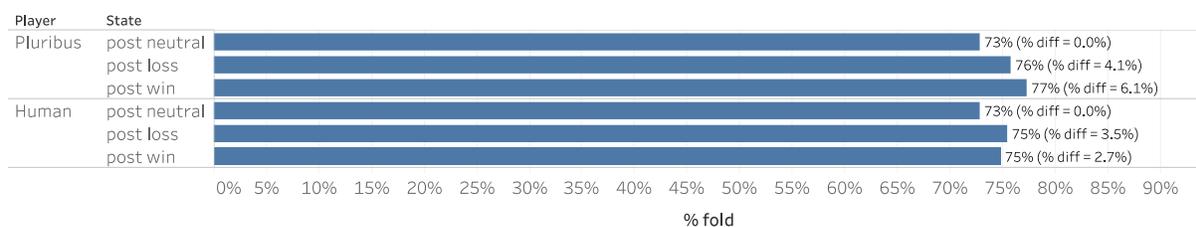

*Figure 5 – Proportion of hands folded by trigger*

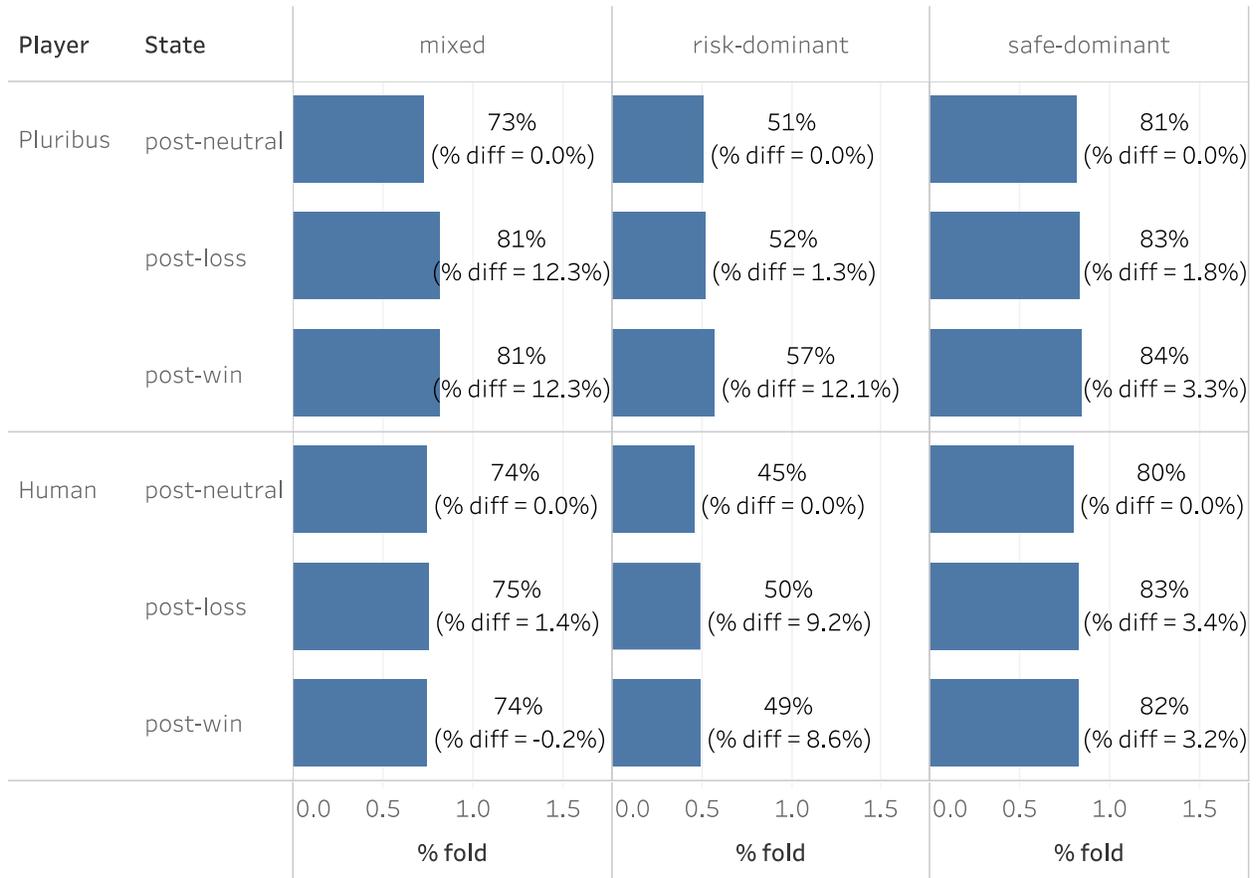

*Figure 6 – Pluribus and the Human increase the number of hands folded following a trigger*

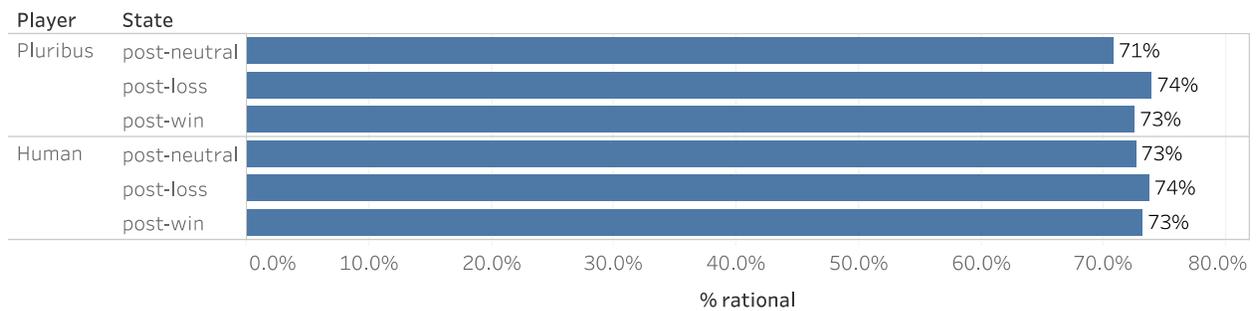

*Figure 7 – Pluribus and the humans increase their aggregate percentage of observed rational decisions following a trigger*

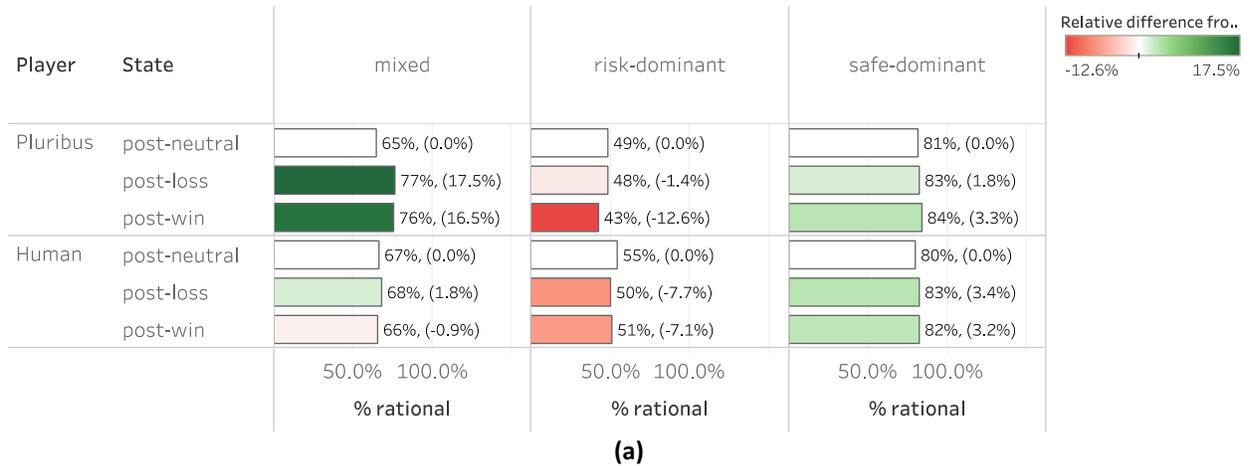

**(a)**

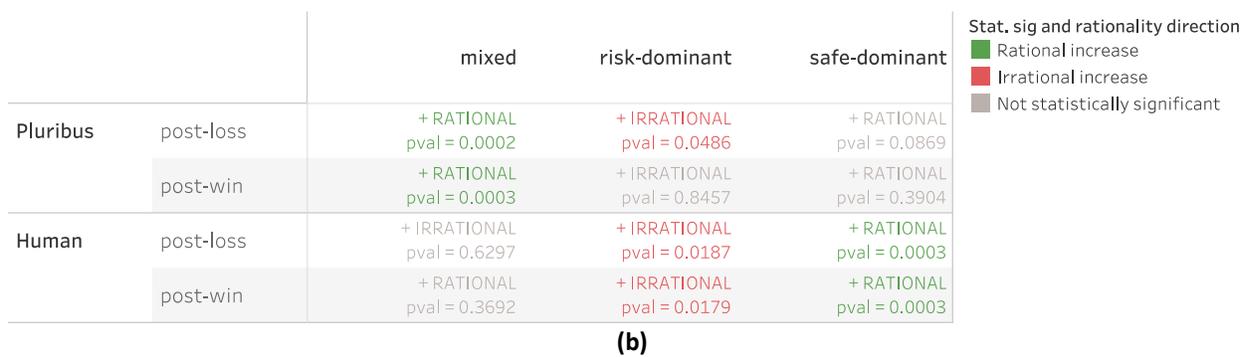

**(b)**

*Figure 8 – Change in proportion of observed rational decisions by gamble type following a trigger (a) and direction / statistical*

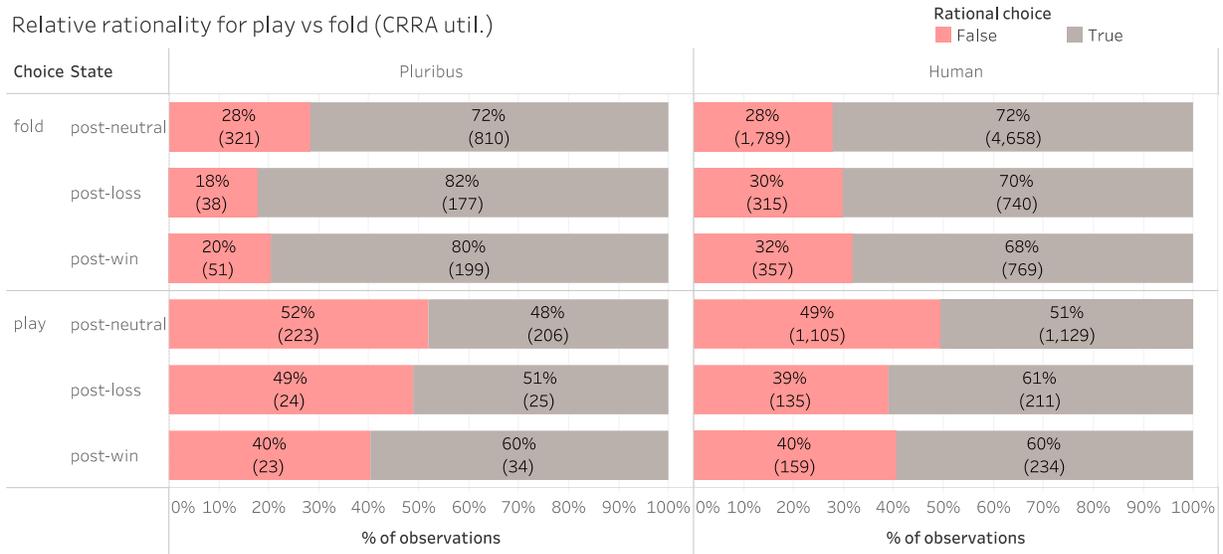

(a)

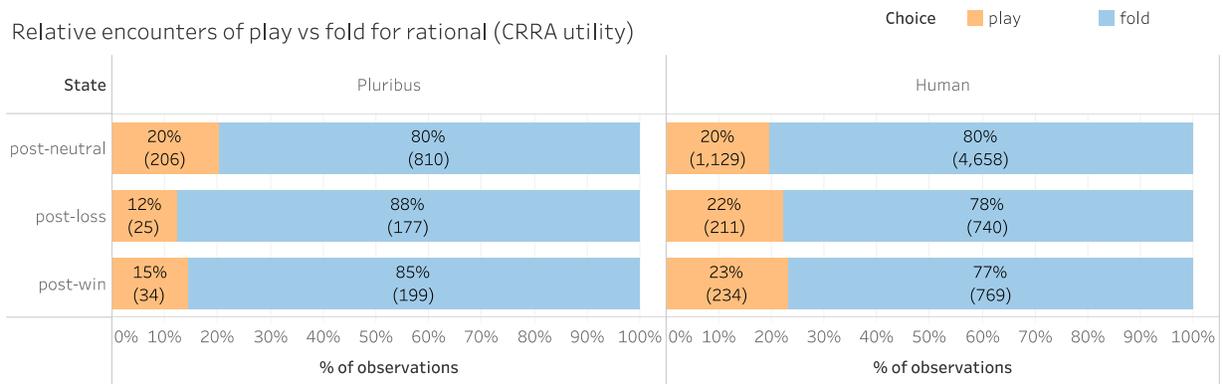

(b)

*Figure 9 - Pluribus increases its proportion of observed rational decisions following a trigger but the Human's shift in observed rationality depends on the type of gamble encountered following a trigger (a). By chance, Pluribus experienced fewer gambles in which the safer option was preferred following a trigger, whereas the Human experienced more.*

## RUM parameter estimates

Pluribus becomes more risk-averse post-loss ($\omega$ increases 2.5x, Figure 10), leading it to fold more hands than it does in the neutral state (8% absolute increase and 12% relative increase) (Figure 6). Post-win it becomes more risk-seeking, but though the change is statistically significant, it is practically negligible. Following a loss Pluribus becomes more rational overall ($\lambda$ increases 6x), but following a win Pluribus becomes more irrational ($\lambda$ decreases ~25% relative decrease, Figure 10). The utilities and RUM probability predictions are not linear in parameters, so the increases in $\omega$ and $\lambda$ should not be interpreted as the corresponding change in the expectation of choosing to play/fold a hand.

Post-loss, Pluribus' increased risk-aversion and observed propensity to fold hands leads to a net increase in rational decisions (Figure 7 and Figure 8), due to the presence of an overwhelming number of hands that should be folded (Figure 9). This results in an "overcorrection" on risky-dominant hands where it folds hands that it should be playing. Post-win, Pluribus becomes more risk-seeking, but the magnitude of this effect is too small to alter its observed rationality over the entire dataset. It continues to fold hands it should have played, particularly for risky-dominant gambles, leading to a decrease in the estimated rationality parameter (Figure 10).

Humans become more risk-seeking ($\omega$ decreases) and more irrational ($\lambda$ decreases) following a loss. For high-uncertainty environments, the increase in risk-seeking leads to an increase in the number of hands the Human plays and (by chance) the proportion of hands that should have been played increases, so these decisions are rational. However, an overwhelming number of hands shouldn't be played, and on these hands the increased propensity toward playing is irrational. On the whole, these effects cancel one another out at the observed-decision level for the mixed gambles. Furthermore, even despite the fact that the Human is likely to choose a risky outcome when they become more risk-seeking in a triggered state, their observed irrationality on risk-dominant gambles increases. The mixed and risk-dominant gambles taken jointly imply that the Human is failing to identify when the safe or risky option is preferred and in a triggered state is even more likely to make this mistake, hence the increase in the irrationality parameter $\lambda$.

Pluribus is more risk-averse than the Human in all states, and the difference is more pronounced in the trigger states (comparisons are statistically significant). Pluribus becomes more risk-averse when triggered but the Human becomes more risk-seeking. Both Pluribus and the Human react more strongly to losses than wins, exhibiting a greater magnitude of change in $\omega$ from the neutral state following a loss. Pluribus generally becomes more rational following a loss and less rational after a win. The Human always becomes less rational under both types of trigger.

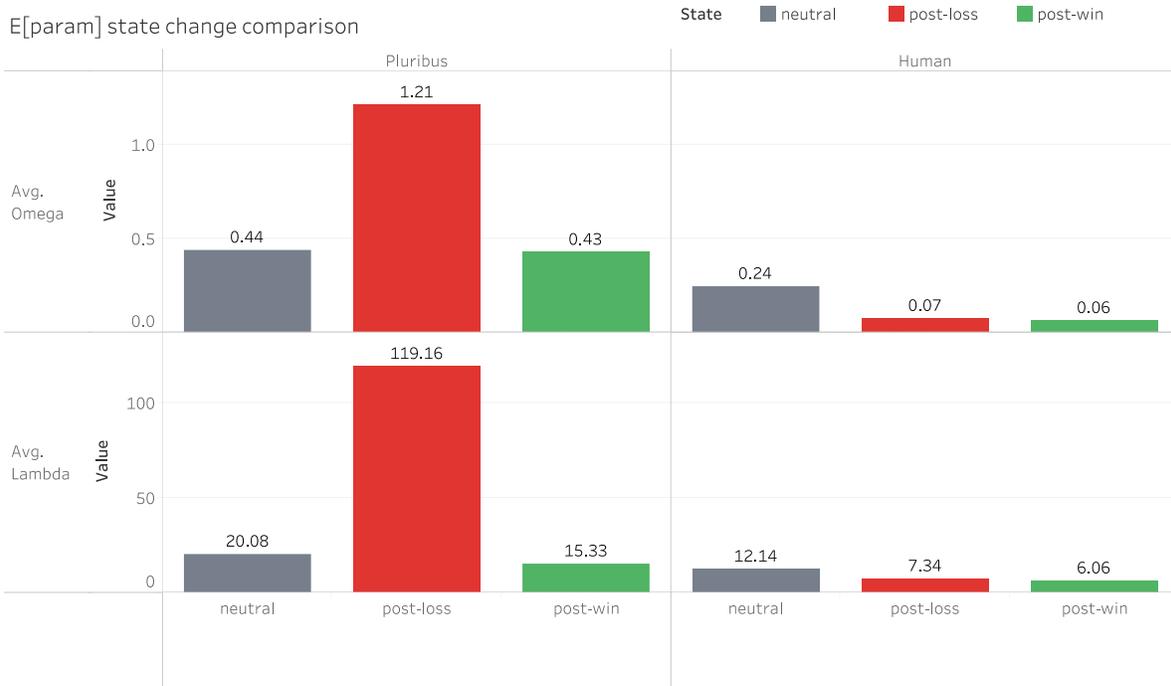

*Figure 10 – Pluribus becomes more risk-averse and more rational following a trigger; the humans become more risk-seeking and irrational*

*Table 3 – Comparison of statistically significant behavior changes by observed outcomes vs. estimated latent parameters*

| | | Observed decisions outcome | | | Random utility model (RUM) | |
|---|---|---|---|---|---|---|
| | | **High-uncertainty** | **Low-uncertainty** | | | |
| | | **Mixed gamble** | **Risk-dominant** | **Safe-dominant** | **Risk-aversion (ω)** | **Rationality (lambda)** |
| **Pluribus** | **Post-loss** | + rational | + irrational | --- | + risk-averse | + rational |
| | **Post-win** | + rational | --- | --- | + risk-seeking | + irrational |
| **Human** | **Post-loss** | --- | + irrational | + rational | + risk-seeking | + irrational |
| | **Post-win** | --- | + irrational | + rational | + risk-seeking | + irrational |

# Discussion

**Hypothesis:** following a trigger event, a human player will exhibit a change in playing style, becoming more or less rational. An algorithmic agent, Pluribus, will not.

We observe a shift in playing style as it relates to rationality for both the Human and Plurib*us* following a loss- or win-trigger event on the two dimensions analyzed: observed decision outcomes and latent RUM risk and rationality parameters!

**Research question 1: does an algorithm change its playing style as a result of a trigger?**

**Yes –** from an observed outcomes perspective, we see Pluribus behave more rationally following a trigger in high-uncertainty environments than it does in a neutral state. The latent model parameters capturing the algorithm's "internal" risk-aversion and rationality also shift following a trigger event. This is contrary to our initial hypothesis. *Ceteris paribus* we wouldn't expect an algorithm to change its behavior in response to what are traditionally thought of as "human feelings" like ego or fear.

### Research question 2: how does playing style change?

Following a trigger event, Pluribus increases its propensity toward folding. This is driven by an increase in risk-aversion and leads to a greater proportion of observed rational decisions on the whole. The RUM rationality parameter indicates that though the observed set of outcomes on the whole may appear more rational, the increase in risk-aversion may mean that there is an increase in irrationality in the algorithm's internal "thought process".

The Human becomes more risk-seeking and more irrational following both loss- and win-trigger events according to the RUM model parameters. In low-uncertainty environments this results in fewer observed rational decisions when presented with risky situations and more rational decisions when presented with safer situations, somewhat counter-intuitively. The Human's ability to integrate the information into high-uncertainty environments is more limited than Pluribus' and we don't see any meaningful change in outcome rationality for the mixed gambles situations.

### Research question 3: why do playing styles change?

For an algorithm, a change in "thought process" could be akin to tuning parameters in its decision-making model. As an example, if our RUM model was identical to the specification of Pluribus' decision-making model – meaning it chose to play or fold according to the CRRA utility function used in this paper – then it could have a dynamic updating procedure such that following a loss or win, the risk-aversion parameter was increased for some number of subsequent decisions, perhaps reflecting a desire to collect more information before putting money at risk following a "surprising" outcome. While we do not have access to Pluribus' full mathematical specification, we know it is built to minimize regret in an inverse reinforcement learning framework [5], so this could be manifesting as risk-aversion in economic frameworks. Returning to the RUM irrationality parameter, regret minimization algorithms may over-compensate in economic terms.

The Human becomes more risk-seeking after a win or loss. Like Pluribus, we can't know why they adjusted their risk-aversion, and they themselves may not know. The behavior exhibited in this dataset does mirror that found by other researchers: humans try to win back money after a loss or get over-confident following a win [8]–[10]. We cannot stress enough, though, that while the shift may be caused by emotion as we suggested as motivation for this research, we cannot know this from this data whether it was emotion or a combination of other factors.

## Conclusions

Humans and algorithms were indistinguishable in their observed decision-making patterns in a neutral state, i.e. in a poker context they played/folded at approximately the same rate on the same quality of hands. However, there was a noticeable differentiation in the number of observed rational decisions, especially in high-uncertainty environments when they experienced a trigger. The economic RUM model is able to partially characterize drivers behind this differentiation: the algorithm became more risk-averse and rational and the Human became more risk-seeking and irrational. This may provide a foundational framework for identifying humans and algorithmic decision-makers in unlabeled contexts: subject the decision-makers to circumstances known to trigger risk tolerance changes in humans and apply RUMs to estimate the change in risk and rationality parameters, labeling the humans and AI according to the types of parameter changes. We are currently working on applying a categorization framework of this nature to this dataset.

In this paper, Pluribus outperforms the Human in high-uncertainty environments in terms of incorporating small amounts of information, or surprise, following a loss or win but may have an unintentional bias toward risk-aversion in an economic context. Behavioral decision-making models offer a possible method for identifying situations in which humans may outperform machines and vice-versa, leading to better integration of the two in joint decision-making contexts.

## Future work

Future work includes estimating similar model structures on other datasets, investigating the accuracy of classifying decisions as human or algorithmic in unlabeled contexts using economic characterization of those decisions, and extending the classification to real-time environments such as online poker games.

# Appendix A
## Deep Neural Net (DNN) and Linear Regression model estimation details

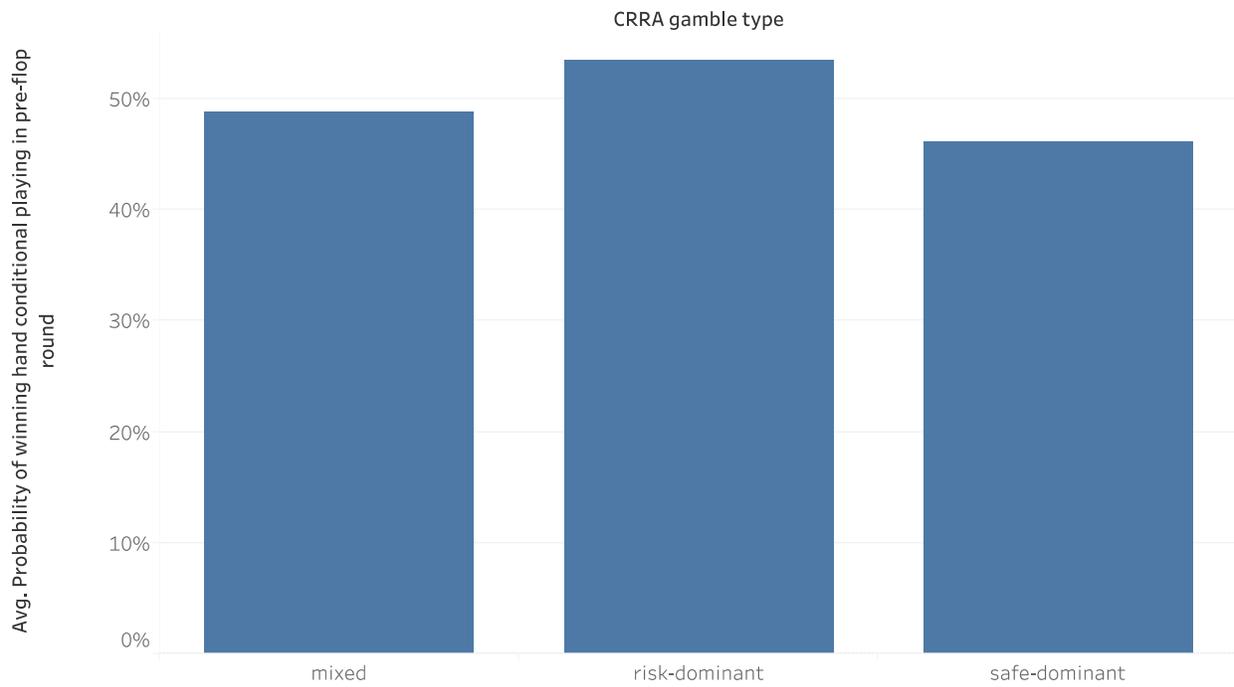

*Figure 11 - Average DNN forecasted probability of winning hand by gamble type*

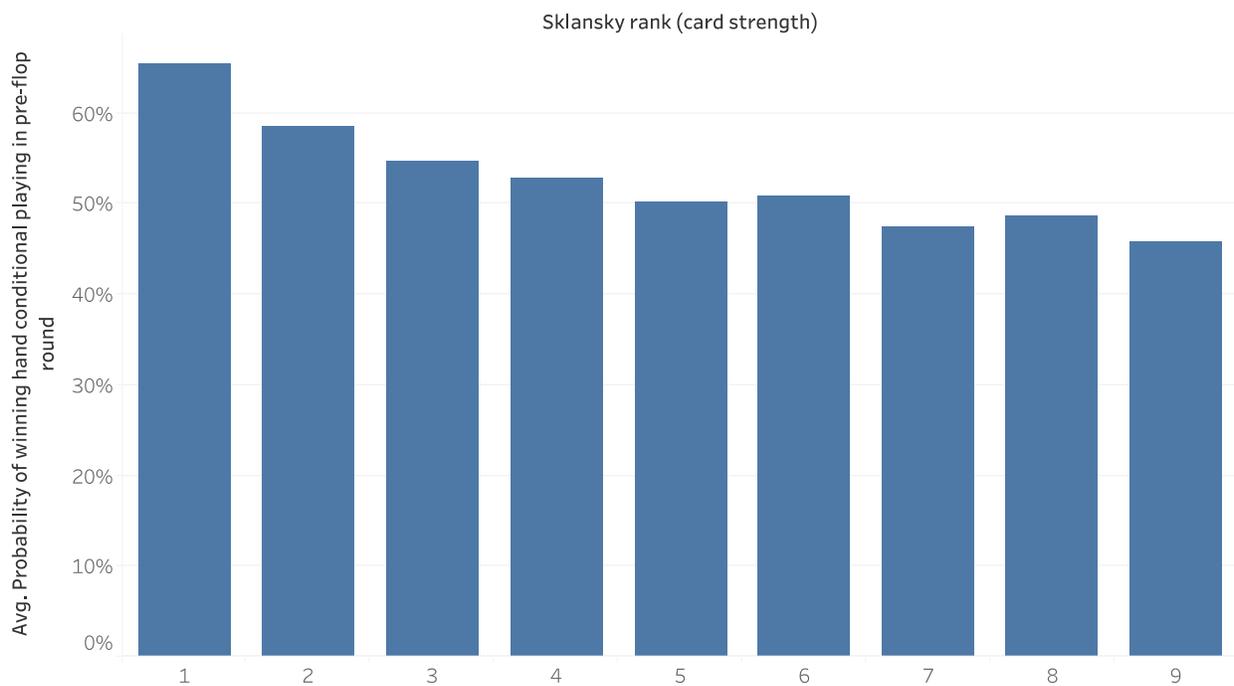

*Figure 12 - Average DNN forecasted probability of winning hand by Sklansky hand strength*

*Table 4 - Difference in expected utilities implied by the deep neural net (DNN) and linear regressions estimated on the dataset to obtain the probabilities and payoffs of each choice situation option*

| | | mixed | risk-dominant | safe-dominant | Grand Total |
|---|---|---|---|---|---|
| Pluribus | post-neutral | 0.0264 (n=1,560) | 0.1098 (n=1,688) | -0.1606 (n=4,301) | **-0.0615** (n=7,549) |
| | post-loss | 0.0188 (n=264) | 0.1110 (n=238) | -0.1418 (n=598) | **-0.0486** (n=1,100) |
| | post-win | 0.0396 (n=307) | 0.1071 (n=299) | -0.1381 (n=678) | **-0.0385** (n=1,284) |
| Human | post-neutral | 0.0204 (n=8,681) | 0.1093 (n=6,167) | -0.1571 (n=22,760) | **-0.0724** (n=37,608) |
| | post-loss | 0.0302 (n=1,401) | 0.1059 (n=913) | -0.1638 (n=3,358) | **-0.0725** (n=5,672) |
| | post-win | 0.0289 (n=1,519) | 0.1095 (n=1,069) | -0.1549 (n=3,797) | **-0.0669** (n=6,385) |

# Appendix B
## Random Utility Model (RUM) model estimation details

*Table 5 - Expected value and sample standard deviation of thirty-five randomly chosen results with valid convergence from multi-start maximum likelihood model estimation. The expected values of these select results were used for evaluation of rational decisions*

|  | post-loss | | post-neutral | | post-win | |
|---|---|---|---|---|---|---|
|  | Human | Pluribus | Human | Pluribus | Human | Pluribus |
| Avg. Omega | 0.073 | 1.213 | 0.243 | 0.435 | 0.061 | 0.428 |
| Std. dev. of omega | 0.0000 | 0.0688 | 0.0002 | 0.0001 | 0.0001 | 0.0001 |
| Avg. Lambda | 7.3 | 119.2 | 12.1 | 20.1 | 6.1 | 15.3 |
| Std. dev. of lambda | 0.0006 | 23.8621 | 0.0074 | 0.0090 | 0.0025 | 0.0060 |

*Table 6 - Statistical significance of between-player parameter differences by state*

| | | |
|---|---|---|
| omega | post-neutral | P1=0.44, P2=0.24<br>pval = 0.0000% |
| | post-loss | P1=1.21, P2=0.07<br>pval = 0.0000% |
| | post-win | P1=0.43, P2=0.06<br>pval = 0.0000% |
| lambda | post-neutral | P1=20.08, P2=12.14<br>pval = 0.0000% |
| | post-loss | P1=119.16, P2=7.34<br>pval = 0.0000% |
| | post-win | P1=15.33, P2=6.06<br>pval = 0.0000% |

*Table 7 - Statistical significance of within-player parameter differences for triggered states as compared to neutral state*

| | | | |
|---|---|---|---|
| **omega** | Human | post loss | P1=0.07, P2=0.24<br>pval = 0.0000% |
| | | post win | P1=0.24, P2=0.06<br>pval = 0.0000% |
| | Pluribus | post loss | P1=1.21, P2=0.44<br>pval = 0.0000% |
| | | post win | P1=0.44, P2=0.43<br>pval = 0.0000% |
| **lambda** | Human | post loss | P1=7.34, P2=12.14<br>pval = 0.0000% |
| | | post win | P1=12.14, P2=6.06<br>pval = 0.0000% |
| | Pluribus | post loss | P1=119.16, P2=20.08<br>pval = 0.0000% |
| | | post win | P1=20.08, P2=15.33<br>pval = 0.0000% |

## Appendix C
### Data file player discrepancies

| Game 1 | Player missing from game 1 that appears in game 2 | Game 2 | Player missing from game 2 that appears in game 1 |
|---|---|---|---|
| 103 | MrOrange | 103b | MrBrown |
| 40 | MrOrange, Hattori, Oren | 40b | MrPink, MrBrown, Eddie |
| 95 | MrOrange | 95b | MrWhite |
| 76 | MrWhite | 76b | ORen |
| 77 | MrWhite | 77b | ORen |
| 94 | MrOrange | 94b | MrWhite |
| 114 | MrBrown | 114b | MrWhite |
| 41 | MrOrange, Hattori, Oren | 41b | MrPink, MrBrown, Eddie |
| 102 | MrOrange | 102b | MrBrown |
| 50 | MrBlue | 50b | MrWhite |
| 113 | MrBrown | 113b | MrWhite |
| 112 | MrBrown | 112b | MrWhite |
| 51 | MrBlue | 51b | MrWhite |
| 44 | MrOrange, Hattori, Oren | 44b | MrPink, MrBrown, Eddie |
| 111 | MrBrown | 111b | MrWhite |
| 52 | MrBlue | 52b | MrWhite |
| 53 | MrBlue | 53b | MrWhite |
| 45 | MrOrange, Hattori, Oren | 45b | MrPink, MrBrown, Eddie |
| 97 | MrOrange | 97b | MrWhite |
| 42 | MrOrange, Hattori, Oren | 42b | MrPink, MrBrown, Eddie |
| 101 | MrOrange | 101b | MrBrown |
| 75 | MrWhite | 75b | ORen |
| 100 | MrOrange | 100b | MrBrown |
| 43 | MrOrange, Hattori, Oren | 43b | MrPink, MrBrown, Eddie |
| 96 | MrOrange | 96b | MrWhite |